          [2001/04/25 1.1 (PWD)]
\documentclass[a4paper,twocolumn]{esapub} 
\usepackage{natbib}
\usepackage{color}
\usepackage{journal}
\usepackage{graphicx}

\title{Evolution of Planets in Disks}
\author{Wilhelm Kley}
\affil{
Institut f\"ur Astronomie \& Astrophysik,
     Abt. Computational Physics,
     Universit\"at T\"ubingen,\\
     Auf der Morgenstelle 10, D-72076 T\"ubingen, Germany
}

\newcommand{\pomega}{{\varpi}}
\def\gapp{\lower 3pt\hbox{${\buildrel > \over \sim}$}\ }
\def\lapp{\lower 3pt\hbox{${\buildrel < \over \sim}$}\ }

\newcommand{\superscr}[1]{^{\rm #1}}

\begin{document}


\maketitle

\begin{abstract}
The main properties of the observed extrasolar planets are reviewed
with respect to their relevance to the formation scenario of
planetary systems. Results of numerical computations of embedded planets
in viscously evolving disks are presented.
Emphasis is given to the accretion and migration process.
New calculations on inviscid disks are shown.

The second part of the talk concentrates on resonant planetary systems.
Among the observed extrasolar systems there are 3 confirmed cases,
Gl 876, HD 82943 and 55 Cnc, where the planets orbit their central
star in a low order mean motion resonance.
Results of numerical simulations modeling the
formation and evolution of such systems are presented.
\end{abstract}
\section{Observations}
Since the first discovery of extrasolar planets around main sequence stars
in 1995, their number has risen to over 100 planets today. 
For an always up-to-date list see
e.g. {\tt {http://www.obspm.fr/encycl/encycl.html}}, 
maintained by J.~Schneider.
Apart from two discoveries by the transit method of OGLE-objects 
\citep{2003Natur.421..507K, 2003A&A...402..791D}, all
others have been found by radial velocity measurements.
Among all systems, there are at least 12 planetary systems with two or more
planets; a summary of their properties has recently been given by
\citet{2002marcy-systems}. Longer observations to study trends
in velocity curves may still increase this number.
Of all stars investigated, planets have been found so far
around about 7 \%.

\begin{figure}
\centering
\includegraphics[width=0.95\linewidth]{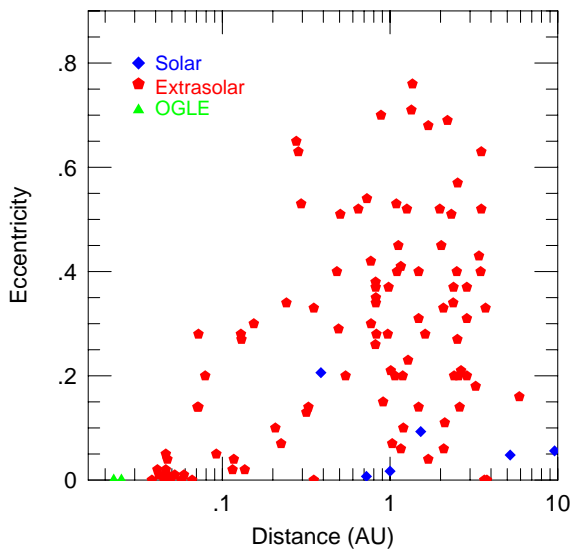}
\caption{Eccentricity vs. distance for the extrasolar
planets. For comparison the solar system planets and the two OGLE
objects are also included.
\label{fig:eccen}}
\end{figure}

The orbital properties of these planetary systems display features
quite different from those of our own solar system. Some statistical
properties are summarized in \citet{2002RvMA...15..133M},
\citet{2003A&A...398..363S} and additional
data may be obtained from {\tt {http://www.obspm.fr/encycl/encycl.html}}.
As an example
we display in Fig.\ref{fig:eccen} the eccentricity versus mass distribution
of all plants detected so far. While the planets close to the star
(hot Jupiters) tend to have circular orbits due to tidal circularization
the planets further out have a rather broad eccentricity distribution
which is essentially indistinguishable from that of binary stars.
Additionally, the cumulative eccentricity-period distribution
of stellar binaries and planetary systems are also nearly
identical for periods
larger than about 10 days \citep{2002RvMA...15..133M}. This
may imply a common origin of the eccentricities, caused for example
by dynamical gravitational interaction.

A summary of the main properties of the exo-planets is outlined
in the following.
The reasons why those characteristics present problems to the standard
planet formation scenarios, is given just below the individual
points, see also \citet{2002Natur.419..355L}.
\begin{list}{I}{~}
\item[{\color{black} $\bullet$}] Large Masses (0.2-17 M$_{\mbox{Jup}}$) \\
Tidal interaction between the planet and the protoplanetary disk
results for planets larger than about 1 Jupiter mass ($M_{Jup}$)
in the creation of a density gap in the disk
\citep{1980MNRAS.191...37L, 1993prpl.conf..749L}. In turn, this gap will
reduce the accretion onto the planet and limit the mass to
about 1 $M_{Jup}$, just as in our solar system
\citep{1993ARA&A..31..129L}.
\item[{\color{black} $\bullet$}] Small Distance (0.04 - 3.3 AU)\\
The high disk temperature in the vicinity of the star makes it difficult
to allow for condensation of solid material. Thus, the cores of
the protoplanets are expected to have been formed beyond the so
called snow line at a distance of about 3 AU and larger from the star
\citep{2000ApJ...528..995S}. 
\item[{\color{black} $\bullet$}] {Large Eccentricities}
 (0.0 - 0.91)  \\
It is usually assumed that planets form in a protoplanetary disk, which
is essentially in a Keplerian rotation around the star. Hence, directly
after formation, planets are expected to orbit on circular orbits.
\item[{\color{black} $\bullet$}] {Resonant Orbits}
 (2:1, 3:1)\\
The condensation process in the disk and the subsequent oligarchic
\citep{1998Icar..131..171K, 2003Icar..161..431T} growth of planetesimals
towards larger planets can occur theoretically at arbitrary orbits. 
For stability reasons, only the planets cannot be on orbits separated
by less than about 10 Hill radii.
\item[{\color{black} $\bullet$}] {Free Floaters}
  (in Jupiter-mass range) \\
Recently, objects in the planetary mass range (a few $M_{Jup}$)
but not orbiting stars have been found
\citep{2000MNRAS.314..858L, 2001ApJ...556..830B}.
Although not directly part of a planetary system today, the origin
of these objects may be related to planet formation, followed by ejection
processes.
It is believed that bodies of a few $M_{Jup}$ are to small
to have formed by direct gravitational collapse.
\end{list}
All the raised features may have their origin in an
evolution process of the young planets still embedded with the
protoplanetary disk from which they formed originally.
\section{Modeling embedded planets}
We shall focus here in the joint evolution
of planets and disks, and concentrate on the fully
non-linear hydrodynamical calculations. 
In recent years
several simulations have been performed
\citep{
1999MNRAS.303..696K, 1999ApJ...514..344B,
1999ApJ...526.1001L, 2000MNRAS.318...18N, 2002A&A...385..647D}.
In the following we describe how these calculation typically
proceed.
\begin{figure}[ht]
\centering
\includegraphics[width=0.95\linewidth]{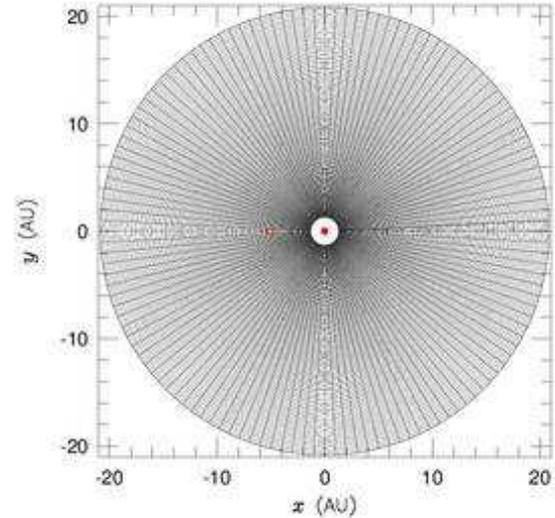}
\caption{A typical computational grid-system ($128 \times 128$)
used in planet-disk computations
\label{fig:grid}}
\end{figure}

All models consider essentially only the late stages of the
planet formation scenario. It is assumed that protoplanets
have already been formed for example by core accretion. The subsequent
evolution of the embedded planets is studied considering the mutual
gravitational interaction between disk and planets. 
In this contribution we shall focus on the purely hydrodynamic
evolution, as elsewhere in this volume Richard Nelson describes magnetic
effects as well.

The ingredients of a model for embedded planets consists of a
star (of typical one solar mass, 1 $M_\odot$), a protoplanetary
disk ($M_{disk} = 0.01 M_\odot$), and a given number of planets.

The disk is modeled by a hydrodynamical evolution in the field
of the star and the planets. 
It is is assumed to be geometrically thin with a small
vertical thickness $H/r << 1$. Mostly it has been modeled by 
a two-dimensional approximation in $r-\varphi$-coordinates
\citep{
1999MNRAS.303..696K, 1999ApJ...514..344B,
1999ApJ...526.1001L, 2000MNRAS.318...18N, 2002A&A...385..647D}.
However recently, also fully three dimensional models
($r, \varphi, \vartheta$) have been performed
\citep{2001ApJ...547..457K, 2003ApJ...586..540D, 2003MNRAS.341..213B}.
To capture all relevant effects global models of the
disk extending typically from about 1 to 30 AU or more have to be
invoked. 
Initial density profiles typically have power laws for the surface
density $\Sigma \propto r^{-s}$ with $s$ between $0.5$ and $1.5$.
Nearly all published models have a fixed radial temperature 
distribution. Mostly, the aspect ratio $H/r$ is held constant in the disk.
From there the temperature profile follows $T(r) \propto r^{-1}$.
In three dimensional models the temperature is vertically constant
and the density $\rho$ follows a Gaussian.

For the anomalous viscosity a Reynolds stress tensor formulation
\citep{1999MNRAS.303..696K}
is used where the kinematic viscosity $\nu$ is either constant
of given by an $\alpha$-prescription $\nu = \alpha c_s H$, where $\alpha$
is constant and $c_s$ is the local sound speed. From observations,
values lying between 0.001 and 0.01 are inferred for the 
$\alpha$-parameter. 
Full MHD-calculations have shown that the viscous stress-tensor ansatz
may give (for sufficiently long time averages)
a reasonable approximation to the mean flow in a turbulent disk
\citep{2003MNRAS.339..983P}.

The embedded planets are assumed to be point masses (using a smoothed
potential), and together with the star they are treated as classical
N-body system. The disk also influences the orbits through the
gravitational torques. This is the desired effect to be studied which
will cause the orbital evolution of the planets.  
The planets may also accrete mass from the surrounding disk.

The initial conditions of the disk are axisymmetric with some given
profile for $\Sigma(r)$ and $T(r)$ as outlined above. Then a Jupiter type
planet is placed into this disk at a distance of several AU (eg. 5.2),
and the joint evolution of the (hydrodynamic) disk and the planetary
(N-body) system is then followed through numerical integration.
In this review we shall not give any details on the numerical issues,
and refer the reader to the appropriate literature
\citep{1998A&A...338L..37K, 1999MNRAS.303..696K, 1999ApJ...514..344B}.
Just to give some impression of how a typical (low-resolution)
grid system looks like we display in Fig.\ref{fig:grid} the
$r-\varphi$ grid having a resolution of $128\times 128$ gridcells.
The radial coordinate extends from 1.3 AU up to 20 AU. For illustration
the star is indicated by central dot and the Roche-lobe of a 1 $M_{Jup}$
planet by the solid line around the planetary position $x_p = -5.2, y_p=0$. 
As can be seen, at this resolution the Roche-lobe of the
planet is merely resolved by a few grid-cells. 
\begin{figure}[t]
\centering
\includegraphics[width=0.95\linewidth]{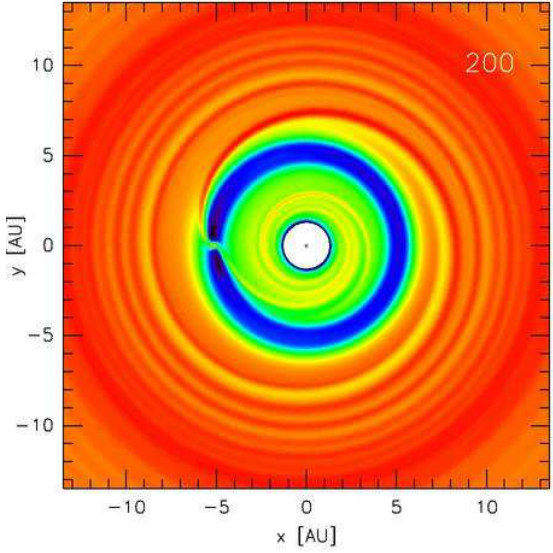}
\caption{Surface density profile for an initially axisymmetric
planet-disk model after 200 hundred orbital orbits of the planet.
\label{fig:pm3-200}}
\end{figure}
\section{Viscous laminar Disks}
The type of modeling outlined in the previous section yields
in general smooth density and velocity profiles, and we refer
to those models as {\it viscous laminar disk} models. In contrast to
models which do not assume an a priori given viscosity and rather
model the turbulent flow directly.
\subsection{The global view}
A typical result of such a viscous computation obtained with
a $128 \times 280$ grid is displayed in
Fig.\ref{fig:pm3-200}. Here, the planet with mass
$M_p = 1 M_{Jup}$ and semi-major axis $a_p = 5.2$AU is 
{\it not} allowed to move and remains on a fixed circular orbit.
The disk has $H/r = 0.01$ and $\nu = 10^{-5}$ in units of
$a_p^2 \Omega_p$, where $\Omega_p$ is the orbital Keplerian
frequency of the planet.
This viscosity for the given temperature is equivalent to
$\alpha = 4 \cdot 10^{-3}$ at the location of the planet.
\begin{figure}
\centering
\includegraphics[width=0.95\linewidth]{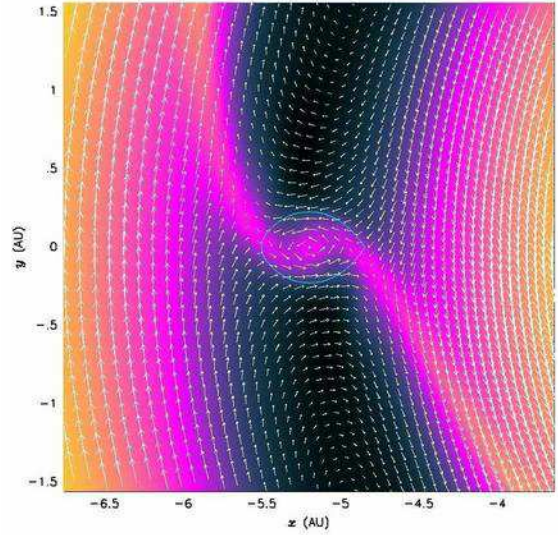}
\caption{Surface density and velocity arrows for a one 
Jupiter mass model.
\label{fig:flow}}
\end{figure}
\begin{figure}
\centering
\includegraphics[width=0.90\linewidth]{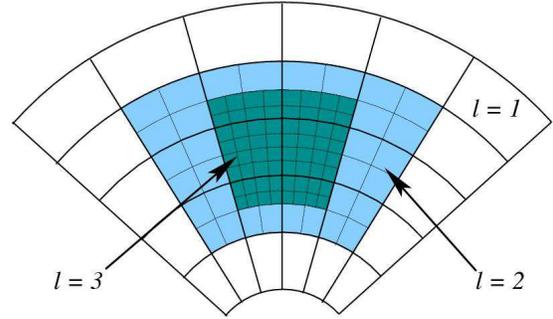}
\caption{Structure of a nested grid centered on the planet. Three
levels of grids are shown (from \citet{2002A&A...385..647D}).
\label{fig:nested}}
\end{figure}
\begin{figure*}[ht]
\resizebox{1.00\textwidth}{!}{%
\includegraphics[clip]{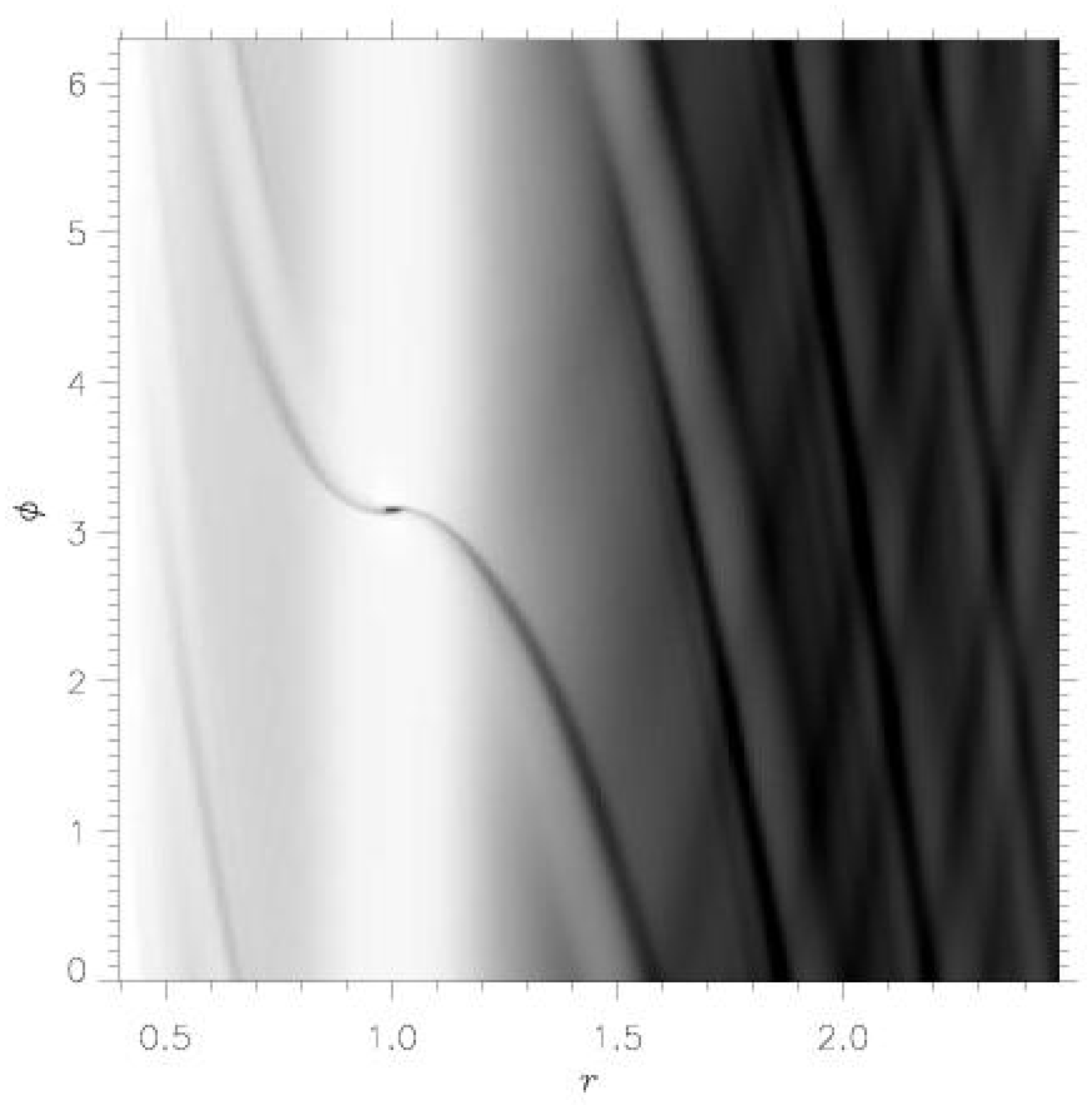}%
\includegraphics[clip]{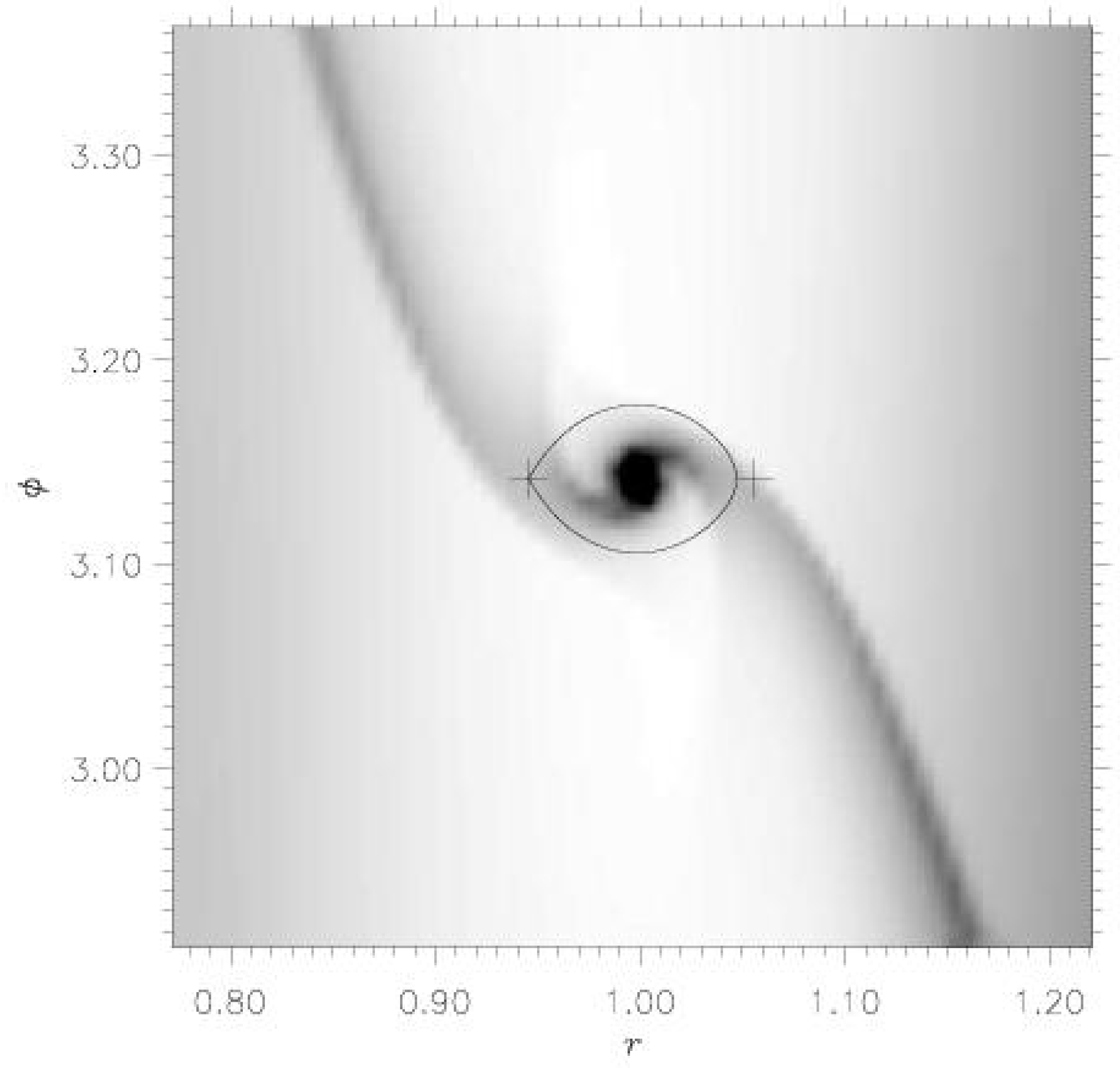}%
\includegraphics[clip]{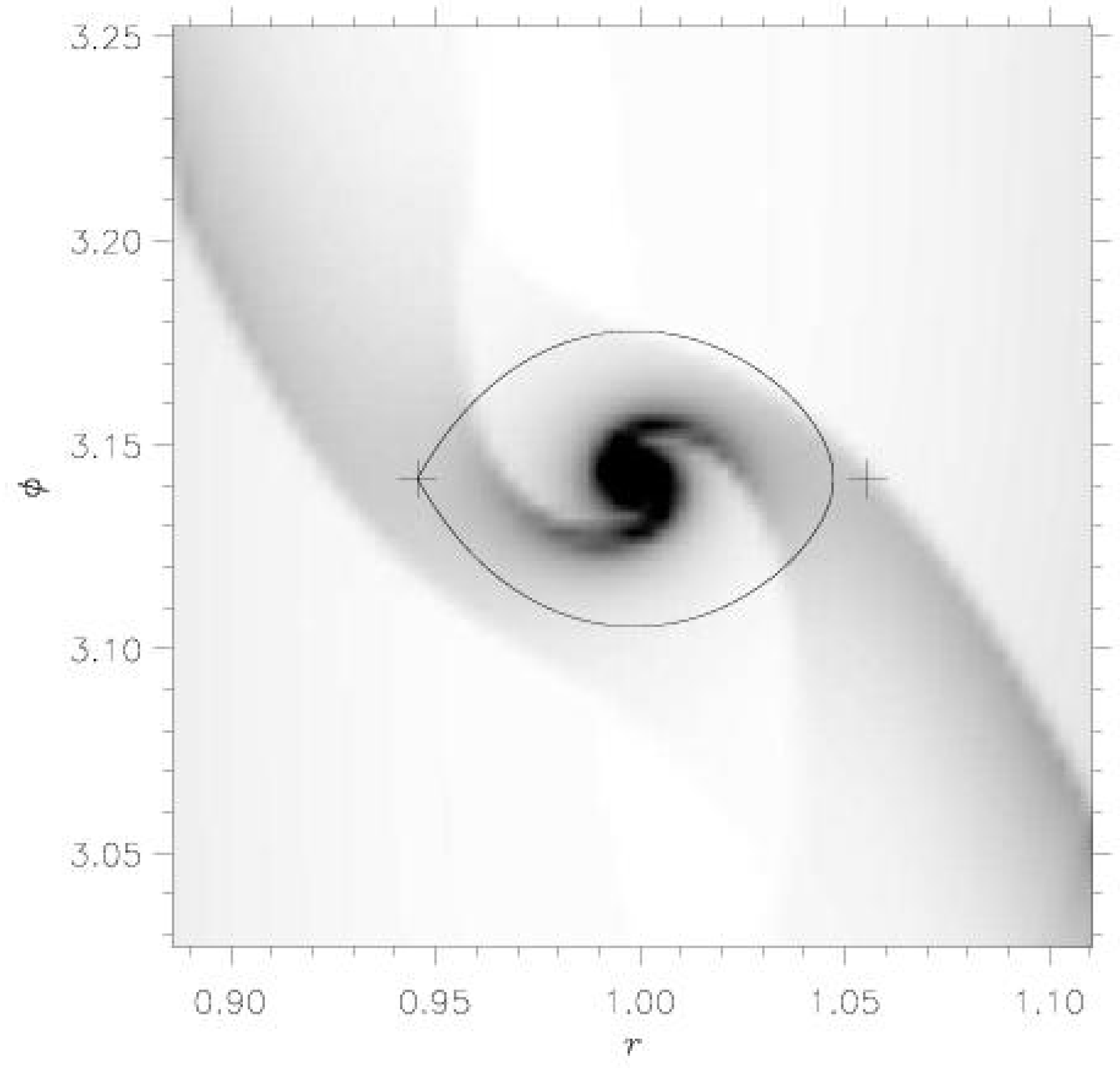}}\\
\resizebox{1.00\textwidth}{!}{%
\includegraphics[clip]{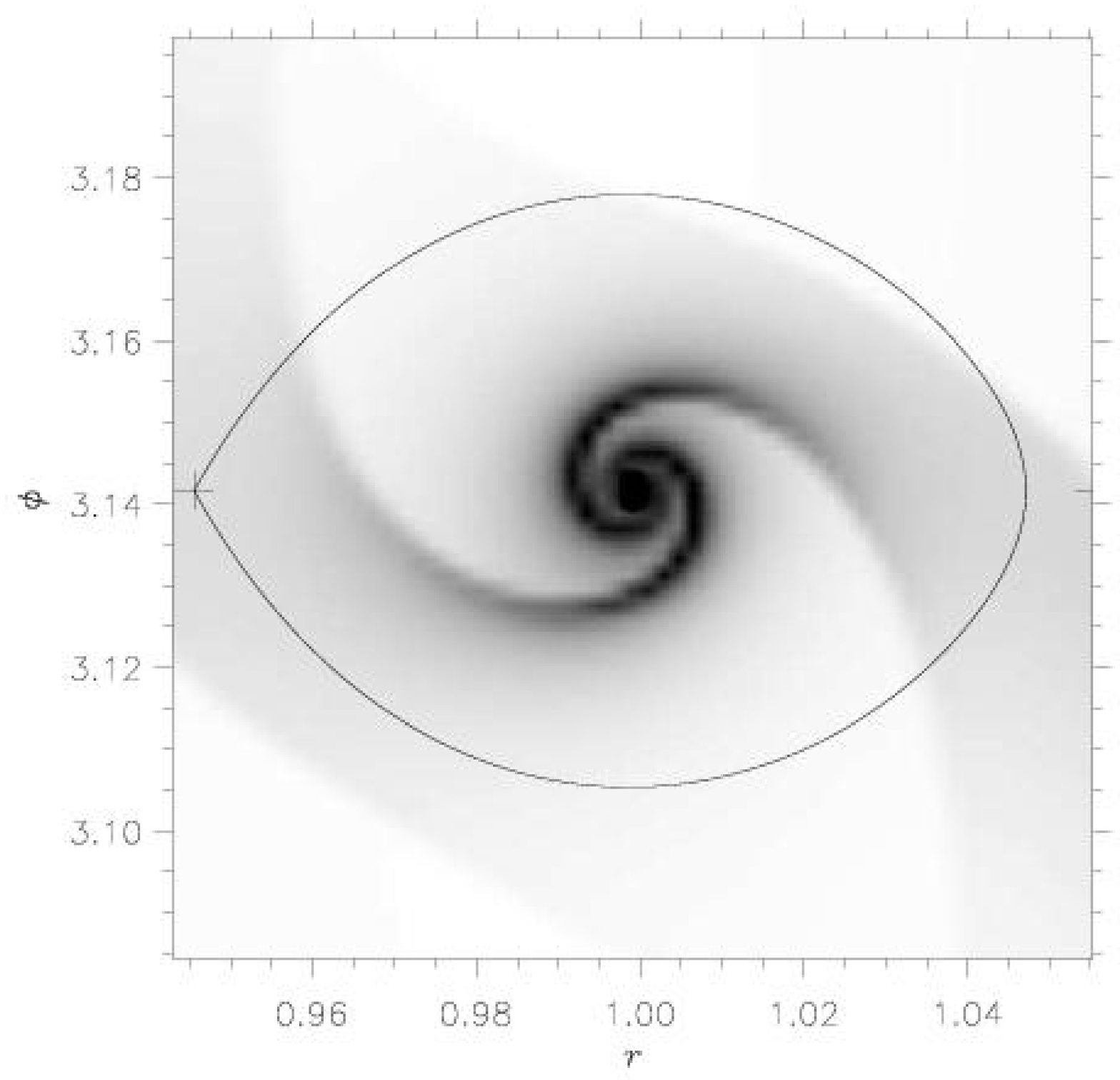}%
\includegraphics[clip]{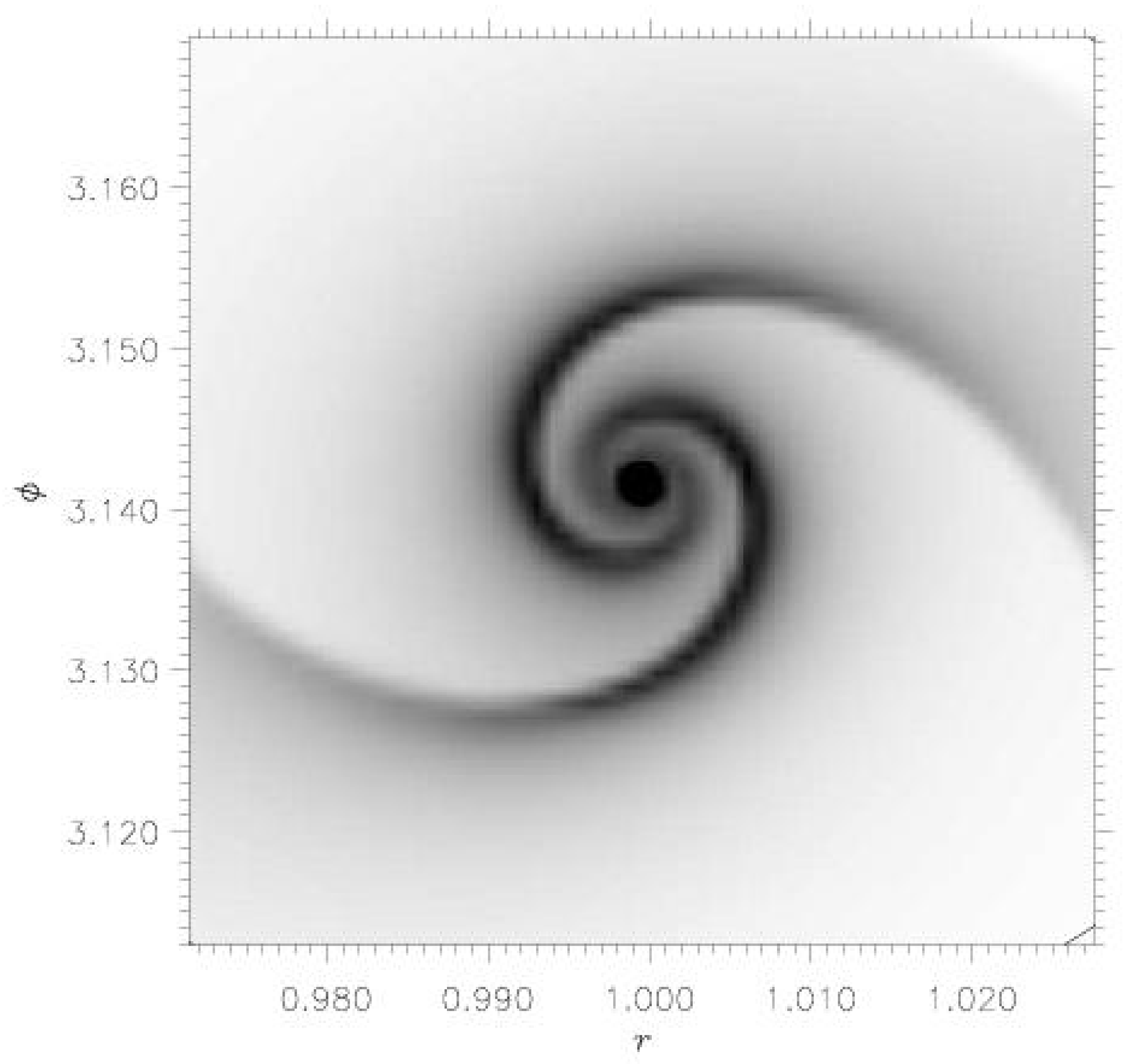}%
\includegraphics[clip]{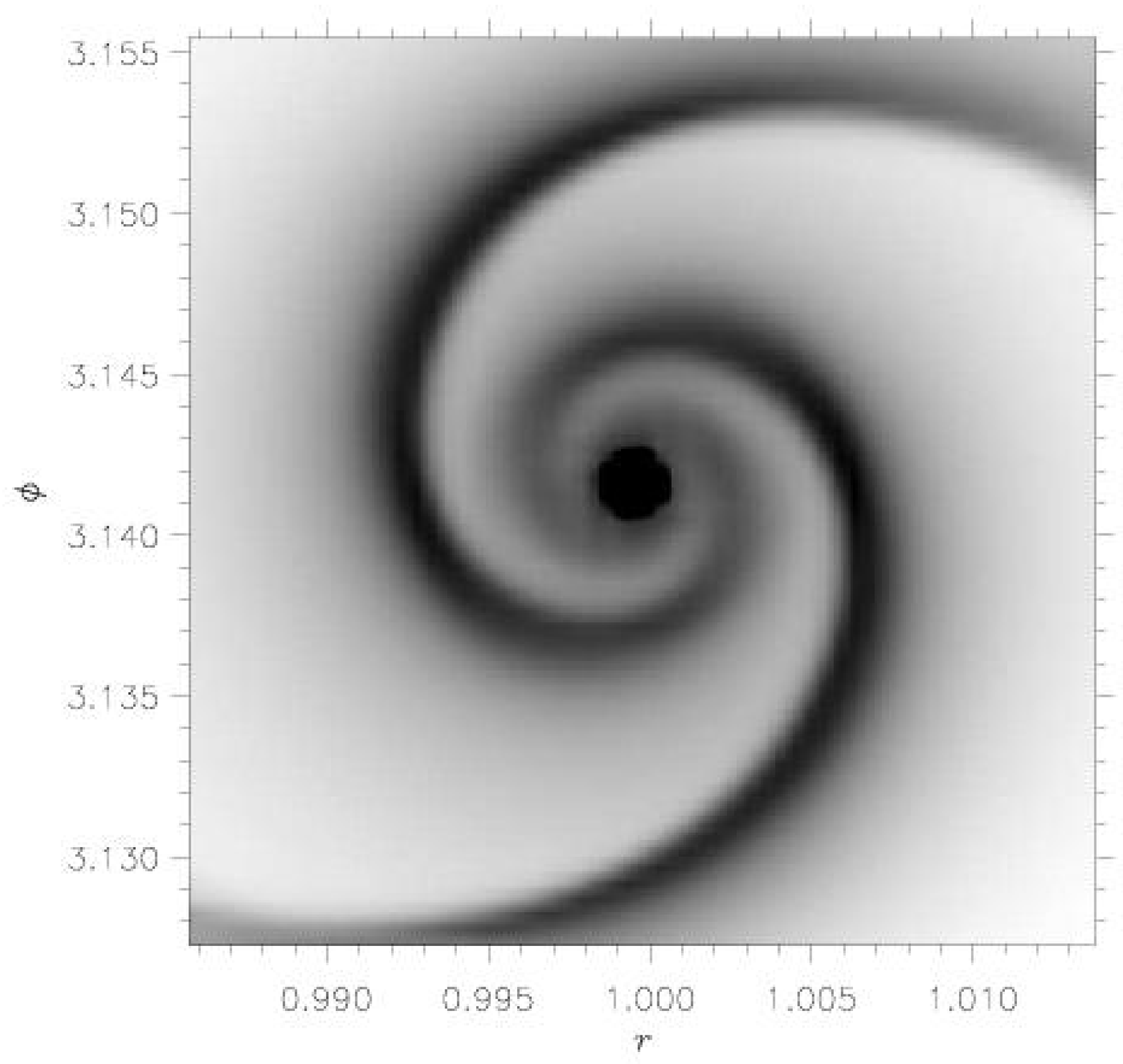}}
\caption{Density Structure of a $1 M_{Jup}$ on each level of the nested
grid system, consisting of 6 grid levels in total.
\label{fig:deep}}
\end{figure*}
Clearly seen are the major effects an embedded planet has on the
protoplanetary accretion disk.
The gravitational force of the planet leads to
spiral wave pattern in the disk. In the present calculation
(Fig.\ref{fig:pm3-200}) there are two spirals in the outer disk
in the inner disk. 
The tightness of the spiral arms depends on
the temperature (i.e. $H/r$) of the disk. The smaller the
temperature the tighter the spirals. The second 
prominent feature is the density gap at the location of the planet.
It is caused by the deposit of positive (at larger radii) and 
negative (at smaller radii) angular momentum in the disk. The spiral
waves are corotating in the frame of the planet, and hence their
pattern speed is faster (outside) and lower (inside) than the 
disk material. Dissipation by shocks or viscosity leads to the
deposit of angular momentum, and pushes material away from the
planet.
The equilibrium width of the gap is determined by the balance
of gap-closing viscous and pressure forces and gap-opening
gravitational torques.
For typical parameter of a protoplanetary disk, a Saturn
mass planet will begin to open a visible gap.

The density structure and flow field 
in the vicinity of the planet for such a model with the
same physical parameter but a resolution of $128 \times 440$ grid-cells,
is displayed in Fig.\ref{fig:flow}.
The solid line represents the size of the Roche-lobe of the
planet. The velocity arrows are calculated with respect to
the planet in the corotating frame.
At this resolution the Roche-lobe is still barely resolved and the
rotation of the material around the planet is visible. It is
noticeable that even for a one Jupiter mass planet with a very deep
gap, mass accretion is still possible. Only beyond about 5 $M_{Jup}$ 
the gap gets to wide and additional accretion is strongly
inhibited.
\subsection{The local view}
To obtain more insight into the flow near the planet and to
calculate accurately the torques of the disk acting on the planet,
a much higher spatial resolution is required. As this is necessary
only in the immediate surrounding of the planet, a number of nested-grid
and also variable grid-size simulations have been performed
\citep{2002A&A...385..647D, 2003ApJ...586..540D, 2003MNRAS.341..213B}.
As an example we display in Fig.\ref{fig:nested} the
structure of such a nested grid system with three levels. The equations
are solved on all levels and appropriate boundary conditions are
formulated to ensure mass, energy and angular momentum conservation
across the grids. This type of grid-system is not adaptive, as
it is defined in the beginning and does not change with time.
The planet is placed in the center of the finest grid.

The result for a 2D computation using 6 grids is displayed in
Fig.\ref{fig:deep},
for more details see also \citet{2002A&A...385..647D}.
The top left base grid has a resolution
of $128 \times 440$ and each sub-grid has a size of $64 \times 64$
with a refinement factor of two from level to level.
It is noticeable that the spiral arms inside the Roche-lobe of the
planet are detached from the global outer spirals. The two-armed spiral
around the planet extends deep inside the Roche-lobe and allow for
the accretion of material onto the planet.
The nested-grid calculations have recently been extended to three dimensions
(3D) and a whole range of planetary masses have been investigated,
starting from 1 Earth mass to a few Jupiter masses
\citep{2003ApJ...586..540D}.
In the 3D case the strength of the spiral arms are weaker and accretion
occurs primarily from regions above and below the midplane of the disk.
\subsection{Accretion and Migration}
\begin{figure}[ht]
\centering
\includegraphics[width=1.00\linewidth]{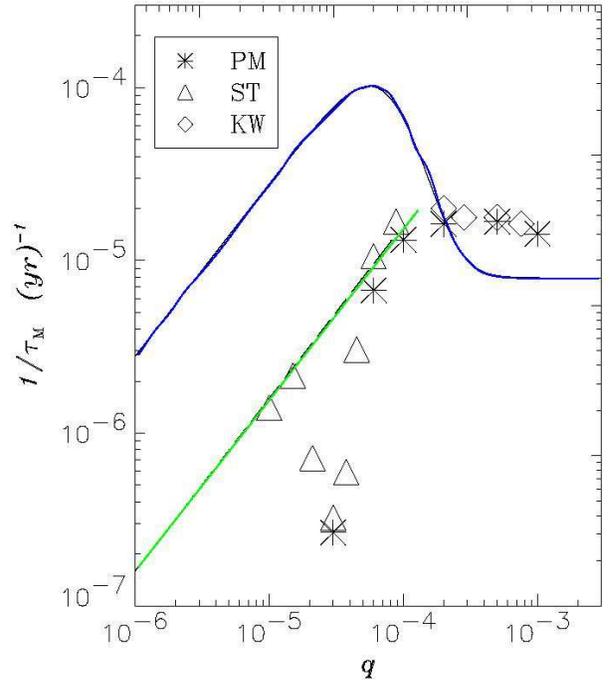}
\caption{The inverse of the migration rate for different planet
masses. The symbols denote different approximations (smoothening) for
the potential of the planet.
\label{fig:mig2}}
\end{figure}

These high-resolution numerical computations allow for a detailed
computation of the torque exerted by the disk material onto the
planet, and its mass accretion rates.
Fig.\ref{fig:mig2} gives the inverse $1/\tau_M$ of the migration 
for three dimensional nested grid calculations as a function of the
planet mass, given in units of the mass of the central star,
$q = M_p/M_\odot$.
The upper dark solid line represents an analytical two-dimensional
linear estimate by \citet{1997Icar..126..261W}
and the lower straight line by \citet{2002ApJ...565.1257T},
which assumes a three-dimensional flow
and takes the corotation torques into account. The symbols refer
to different approximations of the potential of the planet.
It can be seen that for low masses $q \approx 10^{-5}$ and
intermediate masses $q \approx 8 \cdot 10^{-5}$ the 
numerical results fit well to the linear theory. In the intermediate
range of about $q \approx 3 \cdot 10^{-5}$ the migration rates
are about an order of magnitude longer
\citep{2003ApJ...586..540D}. This effect may be caused by
the onset of gap formation in the mass range of about 10-15 earth masses. 
Here nonlinear effects begin to set in and modify the physics. 
These results should be compared to those obtained
by \citet{2003MNRAS.341..213B}.

\begin{figure}
\centering
\includegraphics[width=1.00\linewidth]{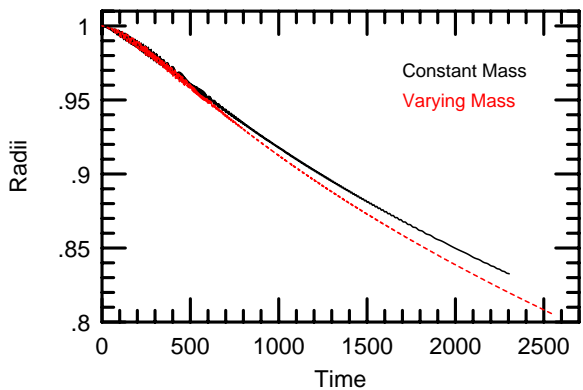}
\caption{The radial position of an initially $1 M_{Jup}$  planet
evolving with the disk. In one of the cases the mass was held fixed
while in the other mass accretion from the disk was taken into account.
\label{fig:r1}}
\end{figure}
The consequences of accretion and migration have been studied by
numerical computations which do not hold the planet
fixed at some radius but rather follow the orbital evolution of the
planet \citep{2000MNRAS.318...18N}.
In Fig.\ref{fig:r1} we display the radial evolution of
an embedded protoplanet as it evolves with the disk.
The typical migrations timescales are found to be of the order
$10^5$~yrs, while the accretion timescale may be slightly smaller.
When planets reach the vicinity of the central star they can reach
up to about 4 $M_{Jup}$.
\section{Inviscid disks}
\begin{figure}
\centering
\includegraphics[width=0.95\linewidth]{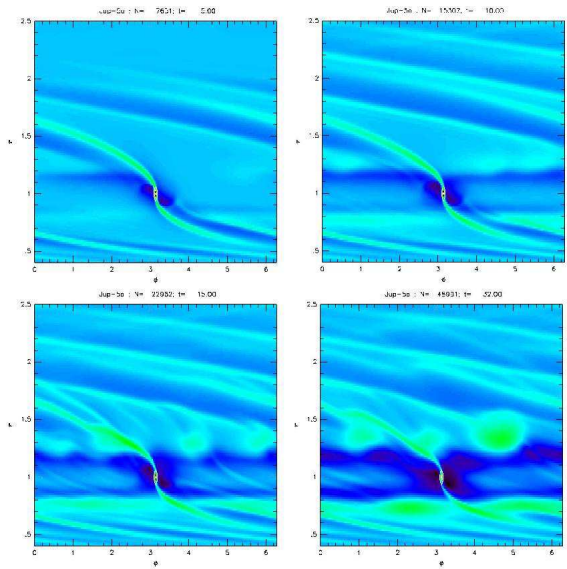}
\caption{Results of an inviscid disk computation with an
embedded 1 $M_{Jup}$ planet. Results are displayed at four different
times (5, 10, 15, and 32) in units of the orbital period of the
planet. Shown are gray-scale plots of the surface density in
an ($r-\varphi$)-coordinate system. 
\label{fig:inviscid}}
\end{figure}
To investigate the influence of viscosity we studied recently
also inviscid models with no physical viscosity added. Only a numerical
bulk viscosity has to be added to ensure numerical stability
\citet{1999MNRAS.303..696K}.
However, this has no influence on the physical effects of the
simulations. In Fig.\ref{fig:inviscid} we display the density
structure of such an inviscid model for a 1 $M_{Jup}$ planet.
The planet is not allowed to move and for stability
purposes its mass is gradually switched on during the first
5 orbits. Results are displayed at 4 different times (5, 10, 15, 32).
After only 5 orbits the spiral waves are already clearly visible,
as they form on a dynamical time scale. The gap is just beginning to
clear. At the edges of the gap high density blobs ({\it vortices}) 
are forming which are reduced in number through merging. Eventually
only one big blob remains. Also inside the gap, some detailed structure
is visible. In comparison, fully viscous simulation show these
vortices mostly as transient features in the beginning of the
simulations. They vanish later on as a result of the viscosity.
If accretion and also migration of the planet is considered, these
moving vortices may create some more complex time dependence.
However, the details of such inviscid models will have to be
studied in future work.
Vortices in accretion disks have sometimes been considered to enhance
planet formation either through triggering a direct gravitational
instability or by trapping particles in it
\citet{2000ApJ...537..396G, 2001MNRAS.323..601D, 2003ApJ...582..869K}.
\section{Resonant systems}
In this part we would like to concentrate on planetary systems which
are in resonance. 
The above mentioned hydrodynamic simulations with single planets
have been extended to models which contain
multiple planets.
It has been shown
\citep{2000MNRAS.313L..47K,2000ApJ...540.1091B,
2001A&A...374.1092S,2002MNRAS.333L..26N}
that during the early evolution,
when the planets are still embedded in the disk,
different migration speeds may lead to an approach of neighboring
planets and eventually to resonant capture.
More specifically, the evolution of planetary systems into a 2:1
resonant configuration was seen in the calculations of
\citet{2000MNRAS.313L..47K}
prior to the discovery of any such systems.

In addition to hydrodynamic disk-planet simulations, many authors have analyzed
the evolution of multiple-planet systems with N-body methods.
Each of the known resonant systems have been considered in detail.
\citet{2002ApJ...572.1041J} and \citet{2002ApJ...567..596L}
have modeled the evolution of 2:1 resonant system GJ~876, while
the 3:1 system 55~Cnc has been analyzed by
\citet{2003ApJ...585L.139J} and \citet{2002astro-ph..0209176},
and the 2:1 system HD 82943 by
\citet{2001ApJ...563L..81G} and \citet{2002astro-ph..0301353}.
Based on orbit integrations,
these papers confirm that the planets in these systems are
in resonance with each other.
The dynamics and stability of resonant planetary
systems in general has been recently studied by
\citet{2002astro-ph..0210577}.

Here we present numerical calculations treating the
evolution of two planets still embedded in a protoplanetary disk.
We use both hydrodynamical simulations and simplified N-body
integrations to follow the evolution of the system. In the first approach,
the disk is evolved by solving the full time-dependent
Navier-Stokes equations simultaneously
with the evolution of the planets. Here, the motion of the planets
is determined by the gravitational action of both planets,
the star, and the disk.
In the latter approach, we take a simplified
approximation and perform 3-body (star plus two planets) calculations
augmented by additional (damping) forces which approximately
account for the gravitational influence of the disk \citep[e.g.][]{2002ApJ...567..596L}.
Using both approaches, allows a direct comparison of the
alternative methods,
and does enable us to determine the damping parameters
required for the simpler (and much faster) second type of approach.

\subsection{Observations}
\label{sec:obs}
The basic orbital parameters of the three known systems in mean motion resonance
are presented in Table~\ref{tab:system}.
The orbital parameters for Gl~876 are taken from the
dynamical fit of \citet{2001ApJ...551L.109L}, and for
HD~82943 from \citet{2001ApJ...563L..81G}.
Due to the uncertainty in the inclinations of the systems,
$M\sin i$, rather than the exact mass of each planet, is listed.
By including the mutual perturbations of the planets into their fit
of Gl~876, \citet{2001ApJ...551L.109L}, however,
are able to constraint that system's inclination to
$\sim 30^{\rm o} - 50\superscr{o}$.

Two of the systems,
Gl~876 and HD~82943, are in a nearly exact 2:1 resonance.
We note that in both cases the outer planet is more massive,
in one case by a factor of about two (HD~82943) and in the
other by more than three (Gl~876).
The eccentricity of the inner (less massive) planet is larger than
that of the outer one in both systems. For the system Gl~876 the alignment of the orbits
is such that the two periastrae are pointing in nearly the same
direction. For the system HD~82943 these data have not been
clearly identified, due to the much longer orbital periods,
but they do not seem to be very different from each other.
The third system, 55~Cnc, is actually a triple system. Here the inner
two planets orbit the star very closely and are in a 3:1
resonance, while the third, most massive planet orbits at a
distance of several AU.
\begin{table}
\caption{
The orbital parameters of the three systems known to contain a mean
motion resonance. $P$ denotes the orbital period,  $M\sin i$ the
mass of the planets, $a$ the semi-major axis, $e$ the eccentricity,
and $\pomega$ the angle of periastron.
It should be noted that the orbital elements for shorter period
planets undergo secular time variations. Thus in principle one should
always state the epoch corresponding to these osculating elements
\citep[see e.g.][]{2001ApJ...551L.109L}.
}

\label{tab:system}
\begin{tabular}{c|l|l|l|l|l}
\hline
   Name  &   P  & $M\sin i$  &  a        &   $e$  &  $\pomega$  \\
       &   [d]  & [$M_{Jup}$]  &  [AU]   &        &   [deg]  \\
 \hline
   Gl~876& (2:1)  &   &   &  &    \\
   c   &  30.1 &   0.56         &  0.13   &   0.24  &  159     \\
   b   &   61.02  & 1.89           &  0.21  &   0.04  &  163     \\
 \hline
   HD~82943 & (2:1)  &   &   &  &   \\
   b   &  221.6  &  0.88          &  0.73   &   0.54 &   138    \\
   c   &  444.6  & 1.63          &  1.16   &   0.41 &    96    \\
 \hline
   55~Cnc & (3:1) &   &   &  &    \\
   b   &  14.65 &  0.84          &  0.11   &   0.02 &   99  \\
   c   &  44.26 &   0.21         &  0.24   &   0.34 &    61   \\
    d   & 5360   & 4.05          &  5.9    &   0.16 &   201    \\
 \hline
\end{tabular}
\end{table}
\subsection{Modeling resonant planets}
\label{sec:model}
The set of coupled hydrodynamical-N-body
models presented in this contribution are calculated in the
same manner as the models described above and in
\citet{1998A&A...338L..37K, 1999MNRAS.303..696K}
for single planets and in \citet{2000MNRAS.313L..47K}
for multiple planets. The reader is referred to those papers
for details on the computational aspects of the simulations.
Other similar models, following explicitly the motion of single and
multiple planets in disks, have been presented by
\citet{2000MNRAS.318...18N}, \citet{2000ApJ...540.1091B},
and \citet{2001A&A...374.1092S}.

The initial hydrodynamic structure of the disk, which extends radially
from $r_{min}$ to $r_{max}$,
is axisymmetric with respect to the location of the star, and
the surface density scales as  $\Sigma(r) = \Sigma_0 \, r^{-1/2}$, with
superimposed initial gaps \citep{2000MNRAS.313L..47K}.
The initial velocity is pure Keplerian rotation ($v_r=0,
v_\varphi = G M_*/r^{1/2}$). Here, we assume a fixed temperature
law with $T(r) \propto r^{-1}$ which follows from the assumed
constant vertical height $H/r = 0.10$.
The kinematic viscosity $\nu$ is parameterized by an
$\alpha$-description $\nu = \alpha c_s H$, with the sound speed
$c_s = H v_\varphi / r$.

In case of the simplified N-body computations, we follow the approach
of \citet{2002ApJ...567..596L} and parameterize the imposed
damping by specifying a
damping rate $a/\dot{a}$ for the outer planet only, as this
is the only planet still in contact with the disk. This value
may depend on time, taking the disk dissipation into account.
The eccentricity damping time scale is a fixed fraction $K$ of the 
semi-major axis damping, i.e. $e/\dot{e} = K a/\dot{a}$.
Thus, in this case only a 3 body system (star and two planets)
is followed by numerical integration, while the effects of the
disk are taking into account simply through imposed additional
forces reflecting the damping of $a$ and $e$.
\subsection{Results}
\label{sec:results}
At the start of the simulations both planets are placed into
an axisymmetric disk at $a_1=4$ and $a_2=10$ AU,
where the density is initialized with partially opened gaps superimposed
on an otherwise smooth radial density profile.
Upon starting the evolution the two main effects are:
\begin{figure}[t]
\begin{center}
\resizebox{0.98\linewidth}{!}{%
\includegraphics{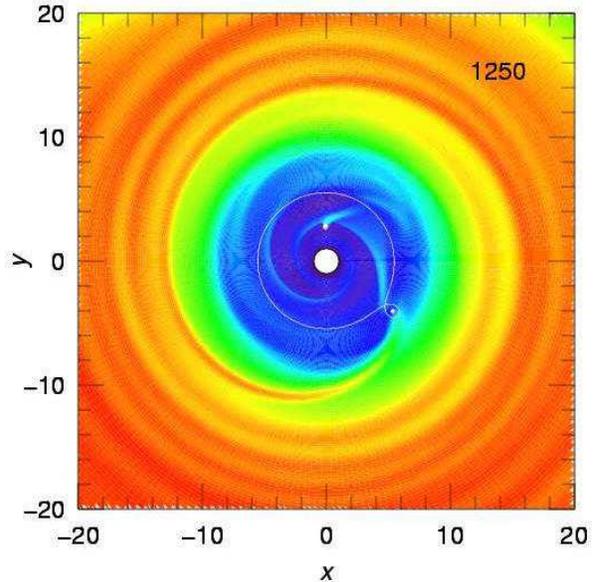}}
\end{center}
  \caption{
  Overview of the density distribution after
  1250 orbital periods of the inner planet.
  Higher density regions are brighter and lower ones are darker.
  The star lies at the center of the white inner region 
  bounded by $r_{min}=1$ AU. The location of the two planets is
  indicated by the
  white dots, and their Roche-lobes are also drawn. Clearly seen are
  the irregular spiral wakes generated by the planets.
  Regular intertwined spiral arms are seen only
  outside of the second planet.
    }
   \label{fig:overview}
\end{figure}
\begin{enumerate}
\item[a)]
Because of the accretion of gas onto the two planets the
radial region in between them is depleted in mass and finally
cleared. This phase typically takes only a few hundred orbital periods.
At the same time the region interior to the inner planet
loses material due to accretion onto the central star.
Thus, after an initial transient phase we typically expect the
configuration of two planets
orbiting within an inner cavity of the disk,
as seen in Fig.\ref{fig:overview}, see also \citet{2000MNRAS.313L..47K}.
\item[b)]
After initialization, the planets quickly (within a few orbital
periods) excite non-axisymmetric disturbances, viz. the spiral waves,
in the disk. In contrast to the single planet case these are
not stationary in time, because there is no preferred rotating
frame.
The gravitational torques exerted on the two planets
by those density perturbations induce a migration process for the planets.
\end{enumerate}
\begin{figure}[ht]
\begin{center}
\resizebox{0.98\linewidth}{!}{%
\includegraphics{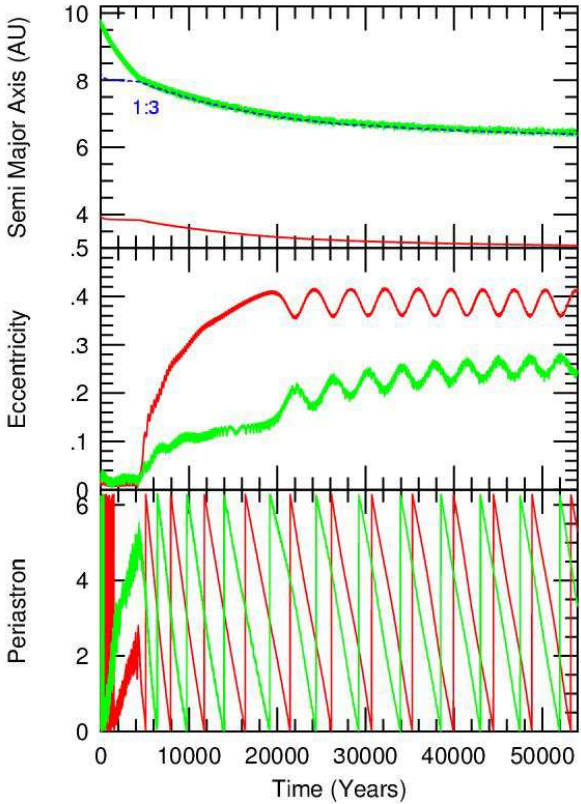}}
\end{center}
  \caption{The semi-major axis ($a$), eccentricity ($e$)
and position angle of the
orbital periastron ($\pomega$) for the two planets versus time.
In this example, the planets have fixed masses of
3 and 5 $M_{Jup}$, and are placed initially at 4 and 10 AU,
respectively. The inner planet is denoted by the black line,
the outer by the light gray line.
The dotted reference line (labeled 3:1), indicates the
location of the 3:1 resonance with respect to the inner
planet.
}
\label{fig:aeo-tw6y}
\end{figure}

Now, the planets' relative positions within the cavity
have a distinct influence on their subsequent evolution.
As a consequence of the clearing process, the inner planet is no longer
surrounded by any disk material and thus cannot grow any further 
in mass. In addition, it cannot migrate anymore, because 
there is no torque-exciting material left in its vicinity.
All the material
of the outer disk is still available, on the other hand, to exert negative
(Lindblad) torques on the outer planet.
Hence, in the initial phase of the computations we observe an
inwardly migrating outer
planet and a stalled inner planet with a constant semi-major axis
(see the first 5000 yrs in the top panel of
Fig.\ref{fig:aeo-tw6y}).

This decrease in separation between the planets increases their gravitational
interactions.
Once the ratio of the planets' orbital periods
has reached a ratio of two integers, i.e. they are close to a
mean motion resonance,
resonant capture of the inner planet by the outer one may ensue.
Whether or not this does actually happen depends on the physical
conditions in the
disk (e.g. viscosity) and the orbital parameters of the planets.
If the migration speed is too large, for example, there may not be enough
time to excite the resonance, and the outer planet will continue
migrating inward \citep[e.g.][]{1999MNRAS.304..185H}.
Also, if the initial eccentricities are too small, then
there may be no capture,
particularly for second-order resonances such as the 3:1 resonance
\citep[see e.g.][]{1999ssd..book.....M}.
\begin{figure}[t]
\begin{center}
\resizebox{0.98\linewidth}{!}{%
\includegraphics{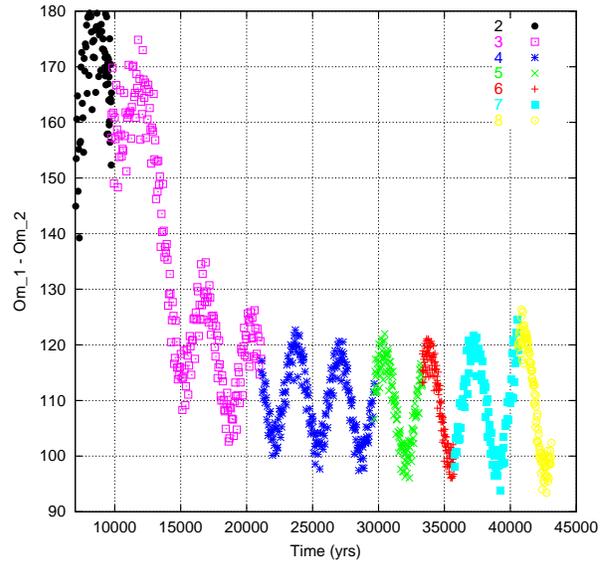}}
\end{center}
\caption{
The evolution of the difference in periastron 
($\Delta \pomega = \pomega_2 - \pomega_1$) vs time for
the two planets. 
\label{fig:o3}
}
\end{figure}
The typical time evolution of the semi-major axis ($a$),
eccentricity ($e$) and direction of the periastron ($\pomega$)
are displayed in Fig.\ref{fig:aeo-tw6y}.
The planets were initialized with zero eccentricities at
distances of 4 and 10 AU in a disk with partially cleared gaps.

In the beginning, after the inner gap has completely cleared,
only the outer planet migrates inward, and the eccentricities
of both planets remain relatively small,
($\lapp 0.02$).
After about 5000 yrs the outer planet has reached a semi-major axis
with an orbital period three times that of the inner planet.
The periodic gravitational forcing leads to
the capture of the inner planet into a 3:1 resonance with the outer one.
This is indicated by the dotted reference line (labeled 3:1)
in the top panel of Fig.\ref{fig:aeo-tw6y}, which
marks the location of the 3:1 resonance with respect to the inner planet.

We summarize the following important features of the evolution
after resonant capture:
\begin{enumerate}
\item[a)]
In the course of the subsequent evolution, the outer planet,
which is still driven inward by the
outer disk material, forces the inner planet to also
migrate inwards.
Both planets migrate inward simultaneously, always retaining their
resonant configuration.
Consequently, the migration speed of the outer planet slows down,
and their radial separation declines.
\item[b)]
Upon resonant capture the eccentricities of both planets grow
initially very fast before settling into an oscillatory
quasi-static state
which changes slowly on a secular time scale.
This slow increase of the eccentricities on the longer time scale
is caused by the growing gravitational forces between the planets,
due to the decreasing radial distance of the two planets
on their inward migration process.
\item[c)]
The ellipses/periastrae of the planets rotate at a constant, retrograde
angular speed $\dot{\pomega}$. Coupled together by the
resonance, the apsidal precession rate $\dot{\pomega}$ for both
planets is identical,
which can be inferred from the parallel lines
in the bottom panel of Fig.\ref{fig:aeo-tw6y}. The orientation
of the orbits is phase-locked with a constant separation
$\Delta \pomega = \pomega_2 - \pomega_1$.
The rotation period of the ellipses (apsidal lines) is slightly longer
than the oscillation period of the eccentricities.
\end{enumerate}
\begin{figure}[t]
\begin{center}
\resizebox{0.98\linewidth}{!}{%
\includegraphics{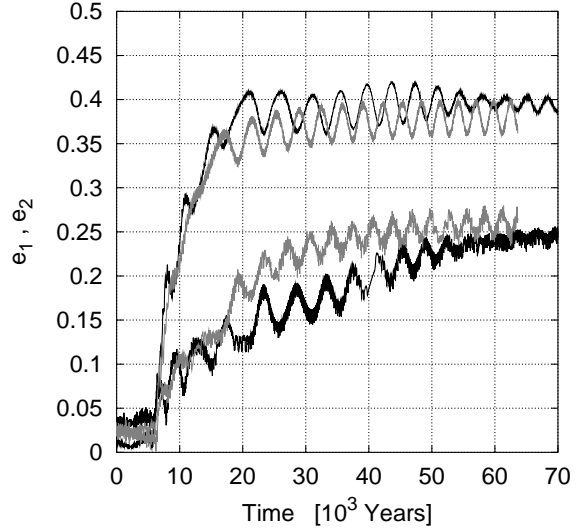}}
\end{center}
\caption{
The evolution of the eccentricities of the inner planet (upper curves)
and the outer planet (lower curves) for the full hydro model
(black curves covering the whole time range) and a simplified 
damped N-body model (light gray), using a damping constant of $K=2.5$.
\label{fig:e7yp}
}
\end{figure}
The capture into resonance and
the subsequent libration of the orbits
is illustrated further in Fig.\ref{fig:o3}.
As suggested in Fig.\ref{fig:aeo-tw6y} (bottom panel)
the periastrae begin to align upon capture in the 3:1 resonance.
Initially, during the phase when the eccentricities are still rising
(between 5 and 20 thousand yrs), the
difference of the periastrae settles intermediately
to $\Delta \pomega \approx 180\superscr{o}$. Then, upon saturation
after about 20,000 yrs, the system re-adjusts and
eventually establishes itself at
$\Delta \pomega \approx 107\superscr{o}$, with a libration amplitude of about
$7\superscr{o}$.
This behaviour can be understood by an analysis of the interaction
Hamiltonian for resonant systems \citep{2002astro-ph..0210577}.
By minimizing the interaction energy, the equilibrium values for
$\Delta \pomega$ (and $\Theta_1$, see below)
can be obtained as a function of the
mass and eccentricity of the two planets.
\subsection{Damped N-body computations}
To illustrate the strengths of a simplified N-body computation
we display in Fig.\ref{fig:e7yp} the eccentricity
evolution of the full hydro and the damped 3-body case.
Despite some differences which we attribute
to the unknown eccentricity damping mechanism, the overall
agreement is reasonable.
For a given semi-major axis damping rate,
the final values obtained for $e_1$ and $e_2$ at larger times depend
on the initial values for the eccentricities
and the amount of eccentricity damping. In this case we used an initial
$e(t_0) = 0.01$ for both planets and an eccentricity damping factor
$K=2.5$, i.e. a slightly shorter
damping time scale for eccentricity as for semi-major axis.
For all models we find that the eccentricity damping rate is of
the same order as the semi-major axis damping, i.e. $K = {\cal{O}}(1)$.
This finding is in contrast to \citet{2002ApJ...567..596L}
who determined a much shorter eccentricity damping time, based
on models for GJ~876.
\begin{figure}[t]
\begin{center}
\resizebox{0.98\linewidth}{!}{%
\includegraphics{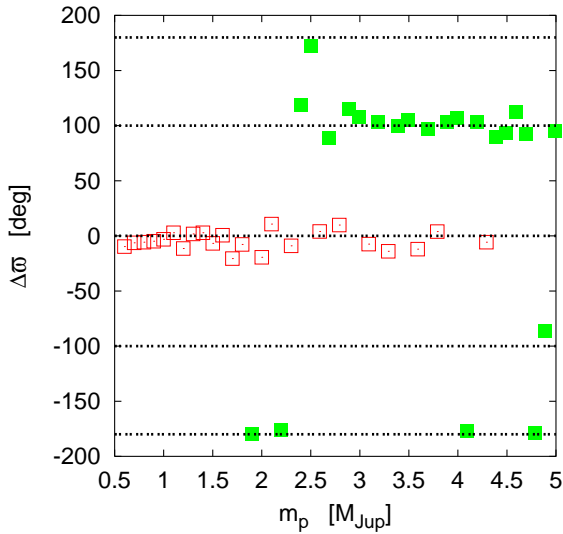}}
\end{center}
\caption{
   Results of a sequence of damped N-body simulations.
   Plotted is the difference of the periastrae
   $\Delta \pomega = \pomega_2 - \pomega_1$ of the two planets after
   capture into resonance versus planet
   mass, where $m_p = m_2 = m_1$.
   The diamonds indicate capture in 2:1 resonance, while the plus
   signs are for 3:1 resonance.
   The other parameters are fixed, as described in the text.
   The horizontal
   lines indicate values of $0$, $\pm 100\superscr{o}$ and 
    $\pm 180\superscr{o}$.
\label{fig:reson12}
}
\end{figure}
\subsection{Resonant angles}
To investigate the mass dependence of the resonant capture process we ran
a sequence of damped N-body models with identical initial conditions
but different masses. 
The results in terms of the resonant angle
$\Delta \pomega = \pomega_2 - \pomega_1$ are displayed in
Fig.\ref{fig:reson12}. 
For small masses $M_p < 2 M_{Jup}$ we find capture occurs only
into the 2:1 resonance. The low masses do not allow for
a sufficiently strong interaction at the 3:1 resonance, and the outer
planet migrates through that point.
All 2:1 resonances settle into the complete symmetric configuration
$\Delta \pomega = \Theta_1 = 0$ (Fig.\ref{fig:reson12}).
For the 2:1 resonance
$\Theta_1 = 2 \lambda_2 - \lambda_1 - \pomega_1$, where $\lambda_i$
denotes the mean longitudes of the planets.

For the 3:1 resonances we see anti-symmetric
configurations with anti-aligned periastrae, 
$|\Delta \pomega | = 180\superscr{o}$ and $\Theta_1 = 0$ for masses
around $m_p = 2 M_{Jup}$, (when the first 3:1 cases begin to occur),
while for all larger masses we find preferentially
the previous non-symmetric configurations, where
$\Delta \pomega \approx 110^{o}$ and $|\Theta_1 | \approx 145\superscr{o}$.
Here, for the case of 3:1 resonance,
$\Theta_1 = 3 \lambda_2 - \lambda_1 - 2 \pomega_1$.
As mentioned above, this behaviour may be an
indication of a bifurcation in the stability properties of resonant
systems, as claimed by \citet{2002astro-ph..0210577}.
With additional damped N-body calculations we also find
that systems entering a 5:2 resonance (not shown)
exhibit a typical anti-symmetric behavior
$|\Delta \pomega | = 180\superscr{o}$
and $\Theta_1 = 0\superscr{o}$.
\section{Discussion}
Through high resolution numerical computations of planets still embedded
in protoplanetary disks it is possible to calculate the
accretion migration rate of planets for the whole mass range from
a few earth masses to over 1 Jupiter mass.
The inferred timescales can be significantly shorter than the
typical lifetimes of the disk.  

Thus, we arrive at the problem of stopping the migration close to the
stars, as well as for Jupiter in our own solar systems.
In the case of the close-in planets it is believed that tidal interaction
with the central star possibly through magnetic fields may be responsible
for braking the planet. In the case of the solar system, early dissipation
of the solar nebula must have occurred to prevent the planets from moving
further in.

Even though a lot of progress has been achieved during the last years,
several problems will have to be addressed in the future. 
Among those are
inclined orbits which may lead to disk warping (see $\beta$ Pic disk),
the inclusion of radiative transport to study observational effects,
planets in binaries to investigate the constraints on the formation.
The study of turbulent (MHD) and radiative
disks has just begun and possibly will lead 
in the future to a change of some aspects of the migration and evolution
scenario of young planets.
%
%
\bibliographystyle{aa}
\bibliography{kley1}

\begin{thebibliography}{43}
\expandafter\ifx\csname natexlab\endcsname\relax\def\natexlab#1{#1}\fi

\bibitem[{{B{\' e}jar} {et~al.}(2001){B{\' e}jar}, {Mart{\'{\i}}n}, {Zapatero
  Osorio}, {Rebolo}, {Barrado y Navascu{\' e}s}, {Bailer-Jones}, {Mundt},
  {Baraffe}, {Chabrier}, \& {Allard}}]{2001ApJ...556..830B}
{B{\' e}jar}, V.~J.~S., {Mart{\'{\i}}n}, E.~L., {Zapatero Osorio}, M.~R.,
  {et~al.} 2001, \apj, 556, 830

\bibitem[{{Bate} {et~al.}(2003){Bate}, {Lubow}, {Ogilvie}, \&
  {Miller}}]{2003MNRAS.341..213B}
{Bate}, M.~R., {Lubow}, S.~H., {Ogilvie}, G.~I., \& {Miller}, K.~A. 2003,
  \mnras, 341, 213

\bibitem[{{Beauge} {et~al.}(2002){Beauge}, {Ferraz-Mello}, \&
  {Michtchenko}}]{2002astro-ph..0210577}
{Beauge}, C., {Ferraz-Mello}, S., \& {Michtchenko}, T.~A. 2002,
  astro-ph/0210577, \apj, submitted

\bibitem[{{Bryden} {et~al.}(1999){Bryden}, {Chen}, {Lin}, {Nelson}, \&
  {Papaloizou}}]{1999ApJ...514..344B}
{Bryden}, G., {Chen}, X., {Lin}, D.~N.~C., {Nelson}, R.~P., \& {Papaloizou},
  J.~C.~B. 1999, \apj, 514, 344

\bibitem[{{Bryden} {et~al.}(2000){Bryden}, {R{\' o}{\. z}yczka}, {Lin}, \&
  {Bodenheimer}}]{2000ApJ...540.1091B}
{Bryden}, G., {R{\' o}{\. z}yczka}, M., {Lin}, D.~N.~C., \& {Bodenheimer}, P.
  2000, \apj, 540, 1091

\bibitem[{{D'Angelo} {et~al.}(2002){D'Angelo}, {Henning}, \&
  {Kley}}]{2002A&A...385..647D}
{D'Angelo}, G., {Henning}, T., \& {Kley}, W. 2002, \aap, 385, 647

\bibitem[{{D'Angelo} {et~al.}(2003){D'Angelo}, {Kley}, \&
  {Henning}}]{2003ApJ...586..540D}
{D'Angelo}, G., {Kley}, W., \& {Henning}, T. 2003, \apj, 586, 540

\bibitem[{{de la Fuente Marcos} \& {Barge}(2001)}]{2001MNRAS.323..601D}
{de la Fuente Marcos}, C. \& {Barge}, P. 2001, \mnras, 323, 601

\bibitem[{{Dreizler} {et~al.}(2003){Dreizler}, {Hauschildt}, {Kley}, {Rauch},
  {Schuh}, {Werner}, \& {Wolff}}]{2003A&A...402..791D}
{Dreizler}, S., {Hauschildt}, P.~H., {Kley}, W., {et~al.} 2003, \aap, 402, 791

\bibitem[{{Go{\' z}dziewski} \& {Maciejewski}(2001)}]{2001ApJ...563L..81G}
{Go{\' z}dziewski}, K. \& {Maciejewski}, A.~J. 2001, \apjl, 563, L81

\bibitem[{{Godon} \& {Livio}(2000)}]{2000ApJ...537..396G}
{Godon}, P. \& {Livio}, M. 2000, \apj, 537, 396

\bibitem[{{Haghighipour}(1999)}]{1999MNRAS.304..185H}
{Haghighipour}, N. 1999, \mnras, 304, 185

\bibitem[{{Ji} {et~al.}(2003{\natexlab{a}}){Ji}, {Kinoshita}, {Liu}, {Guangyu},
  \& {Nakai}}]{2002astro-ph..0301353}
{Ji}, J., {Kinoshita}, H., {Liu}, L., {Guangyu}, L., \& {Nakai}, H.
  2003{\natexlab{a}}, in Proceedings of IAU 189 Colloquium, Sept. 2002,
  Nanjing, P.R. China, in press, astro--ph/0301353

\bibitem[{{Ji} {et~al.}(2003{\natexlab{b}}){Ji}, {Kinoshita}, {Liu}, \&
  {Li}}]{2003ApJ...585L.139J}
{Ji}, J., {Kinoshita}, H., {Liu}, L., \& {Li}, G. 2003{\natexlab{b}}, \apjl,
  585, L139

\bibitem[{{Ji} {et~al.}(2002){Ji}, {Li}, \& {Liu}}]{2002ApJ...572.1041J}
{Ji}, J., {Li}, G., \& {Liu}, L. 2002, \apj, 572, 1041

\bibitem[{{Klahr} \& {Bodenheimer}(2003)}]{2003ApJ...582..869K}
{Klahr}, H.~H. \& {Bodenheimer}, P. 2003, \apj, 582, 869

\bibitem[{{Kley}(1998)}]{1998A&A...338L..37K}
{Kley}, W. 1998, \aap, 338, L37

\bibitem[{{Kley}(1999)}]{1999MNRAS.303..696K}
---. 1999, \mnras, 303, 696

\bibitem[{{Kley}(2000)}]{2000MNRAS.313L..47K}
---. 2000, \mnras, 313, L47

\bibitem[{{Kley} {et~al.}(2001){Kley}, {D'Angelo}, \&
  {Henning}}]{2001ApJ...547..457K}
{Kley}, W., {D'Angelo}, G., \& {Henning}, T. 2001, \apj, 547, 457

\bibitem[{{Kokubo} \& {Ida}(1998)}]{1998Icar..131..171K}
{Kokubo}, E. \& {Ida}, S. 1998, Icarus, 131, 171

\bibitem[{{Konacki} {et~al.}(2003){Konacki}, {Torres}, {Jha}, \&
  {Sasselov}}]{2003Natur.421..507K}
{Konacki}, M., {Torres}, G., {Jha}, S., \& {Sasselov}, D.~D. 2003, \nat, 421,
  507

\bibitem[{{Laughlin} \& {Chambers}(2001)}]{2001ApJ...551L.109L}
{Laughlin}, G. \& {Chambers}, J.~E. 2001, \apjl, 551, L109

\bibitem[{{Lee} \& {Peale}(2002{\natexlab{a}})}]{2002ApJ...567..596L}
{Lee}, M.~H. \& {Peale}, S.~J. 2002{\natexlab{a}}, \apj, 567, 596

\bibitem[{{Lee} \& {Peale}(2002{\natexlab{b}})}]{2002astro-ph..0209176}
{Lee}, M.~H. \& {Peale}, S.~J. 2002{\natexlab{b}}, in Scientific Frontiers in
  Research on Extrasolar Planets, ASP Conference Series, in press,
  astro--ph/0209176

\bibitem[{{Lin} \& {Papaloizou}(1980)}]{1980MNRAS.191...37L}
{Lin}, D.~N.~C. \& {Papaloizou}, J. 1980, \mnras, 191, 37

\bibitem[{{Lin} \& {Papaloizou}(1993)}]{1993prpl.conf..749L}
{Lin}, D.~N.~C. \& {Papaloizou}, J.~C.~B. 1993, in Protostars and Planets III,
  749--835

\bibitem[{{Lissauer}(1993)}]{1993ARA&A..31..129L}
{Lissauer}, J.~J. 1993, \araa, 31, 129

\bibitem[{{Lissauer}(2002)}]{2002Natur.419..355L}
---. 2002, \nat, 419, 355

\bibitem[{{Lubow} {et~al.}(1999){Lubow}, {Seibert}, \&
  {Artymowicz}}]{1999ApJ...526.1001L}
{Lubow}, S.~H., {Seibert}, M., \& {Artymowicz}, P. 1999, \apj, 526, 1001

\bibitem[{{Lucas} \& {Roche}(2000)}]{2000MNRAS.314..858L}
{Lucas}, P.~W. \& {Roche}, P.~F. 2000, \mnras, 314, 858

\bibitem[{{Marcy} {et~al.}(2003){Marcy}, {Fischer}, , {Butler}, \&
  {Vogt}}]{2002marcy-systems}
{Marcy}, G.~W., {Fischer}, D.~A., , {Butler}, R.~P., \& {Vogt}, S.~S. 2003, in
  Space Science Reviews, in press

\bibitem[{{Mazeh} \& {Zucker}(2002)}]{2002RvMA...15..133M}
{Mazeh}, T. \& {Zucker}, S. 2002, Reviews of Modern Astronomy, 15, 133

\bibitem[{{Murray} \& {Dermott}(1999)}]{1999ssd..book.....M}
{Murray}, C.~D. \& {Dermott}, S.~F. 1999, {Solar system dynamics} (Solar system
  dynamics by Murray, C.~D., 1999)

\bibitem[{{Nelson} \& {Papaloizou}(2002)}]{2002MNRAS.333L..26N}
{Nelson}, R.~P. \& {Papaloizou}, J.~C.~B. 2002, \mnras, 333, L26

\bibitem[{{Nelson} {et~al.}(2000){Nelson}, {Papaloizou}, {Masset}, \&
  {Kley}}]{2000MNRAS.318...18N}
{Nelson}, R.~P., {Papaloizou}, J.~C.~B., {Masset}, F.~.~., \& {Kley}, W. 2000,
  \mnras, 318, 18

\bibitem[{{Papaloizou} \& {Nelson}(2003)}]{2003MNRAS.339..983P}
{Papaloizou}, J.~C.~B. \& {Nelson}, R.~P. 2003, \mnras, 339, 983

\bibitem[{{Santos} {et~al.}(2003){Santos}, {Israelian}, {Mayor}, {Rebolo}, \&
  {Udry}}]{2003A&A...398..363S}
{Santos}, N.~C., {Israelian}, G., {Mayor}, M., {Rebolo}, R., \& {Udry}, S.
  2003, \aap, 398, 363

\bibitem[{{Sasselov} \& {Lecar}(2000)}]{2000ApJ...528..995S}
{Sasselov}, D.~D. \& {Lecar}, M. 2000, \apj, 528, 995

\bibitem[{{Snellgrove} {et~al.}(2001){Snellgrove}, {Papaloizou}, \&
  {Nelson}}]{2001A&A...374.1092S}
{Snellgrove}, M.~D., {Papaloizou}, J.~C.~B., \& {Nelson}, R.~P. 2001, \aap,
  374, 1092

\bibitem[{{Tanaka} {et~al.}(2002){Tanaka}, {Takeuchi}, \&
  {Ward}}]{2002ApJ...565.1257T}
{Tanaka}, H., {Takeuchi}, T., \& {Ward}, W.~R. 2002, \apj, 565, 1257

\bibitem[{{Thommes} {et~al.}(2003){Thommes}, {Duncan}, \&
  {Levison}}]{2003Icar..161..431T}
{Thommes}, E.~W., {Duncan}, M.~J., \& {Levison}, H.~F. 2003, Icarus, 161, 431

\bibitem[{{Ward}(1997)}]{1997Icar..126..261W}
{Ward}, W.~R. 1997, Icarus, 126, 261

\end{thebibliography}
\end{document}